\documentclass[10pt,a4paper]{article}
\usepackage[utf8]{inputenc}
\usepackage{amsmath}
\usepackage{amssymb}
\usepackage{graphicx}
\usepackage[margin=1in]{geometry}
\usepackage[labelfont=bf]{caption}

\newcommand{\ket}[1]{\lvert #1 \rangle}
\newcommand{\nfifteen}{$^{15}$N}
\newcommand{\nfourteen}{$^{14}$N}

\newcommand{\freqsensitivity}{13}
\newcommand{\chargesensitivity}{5.3}
\newcommand{\voltagesensitivity}{0.31}

\begin{document}
	
	\title{Nanoscale vector electric field imaging using a single electron spin}
	\author{M.S.J Barson$^1$, L.M. Oberg$^1$, L.P. McGuinness$^1$, A. Denisenko$^2$, N.B. Manson$^1$, J. Wrachtrup$^2$,  \\ and M.W. Doherty$^{1*}$}
\date{$^1$ Laser Physics Centre, Research School of Physics, Australian National University, Acton, ACT 2601, Australia\\
	$^2$ 3. Physikalisches Institut, University of Stuttgart, Pfaffenwaldring 57, 70569 Stuttgart, Germany \\
	$^*$ Corresponding author. Email: marcus.doherty@anu.edu.au}
\maketitle

\begin{abstract}
	The ability to perform nanoscale electric field imaging of elementary charges at ambient temperatures will have diverse interdisciplinary applications. While the nitrogen-vacancy (NV) center in diamond is capable of high-sensitivity electrometry, demonstrations have so far been limited to macroscopic field features or detection of single charges internal to diamond itself. In this work we greatly extend these capabilities by using a shallow NV center to image the electric field of a charged atomic force microscope tip with nanoscale resolution. This is achieved by measuring Stark shifts in the NV spin-resonance due to AC electric fields. To achieve this feat we employ for the first time, the integration of Qdyne with scanning quantum microscopy. We demonstrate near single charge sensitivity of $\eta_e = 5.3$ charges/$\sqrt{\text{Hz}}$, and sub-charge detection ($0.68e$). This proof-of-concept experiment provides the motivation for further sensing and imaging of electric fields using NV centers in diamond.
\end{abstract}

Electrical phenomena are ubiquitous within the physical sciences. The ability to image electric fields is therefore fundamentally important with diverse interdisciplinary applications. Consequently, a myriad of techniques have been developed for precision electrometry. These have demonstrated nanoscale spatial resolution \cite{Yoo579,Martin2008,doi:10.1063/1.358819,doi:10.1063/1.102312}, elementary charge detection \cite{Yoo579,Martin2008,Devoret2000,PhysRevLett.65.3162,doi:10.1063/1.99224,Cleland1998,Bunch490,Salfi2010,Lee_2008}, and the ability to operate at ambient temperatures and pressures \cite{Bunch490,Lee_2008}. However, no device currently possesses all three of these properties simultaneously. A device capable of nanoscale imaging of elementary charges at room temperature would offer unparalleled insight into a variety of systems previously inaccessible to existing electrometers. For example, as a critical characterization tool for two-dimensional electronics. These materials promise a flexible low-energy electronic revolution. Since they are only atomically thick, high-precision field-imaging of single electrons/holes at ambient conditions is required to provide insight into their performance and function, but their charge transport properties are often strongly temperature dependent \cite{doi:10.1021/nn202852j,Novoselov666}. Enhanced electrometers may also have applications in imaging chemical processes (e.g., photosynthesis) and charge phenomena in biological systems (e.g., neuron firing \cite{10.1117/1.NPh.7.3.035002}) that are only relevant at ambient conditions.

\begin{figure}
	\centering
	\includegraphics[width=0.9\textwidth]{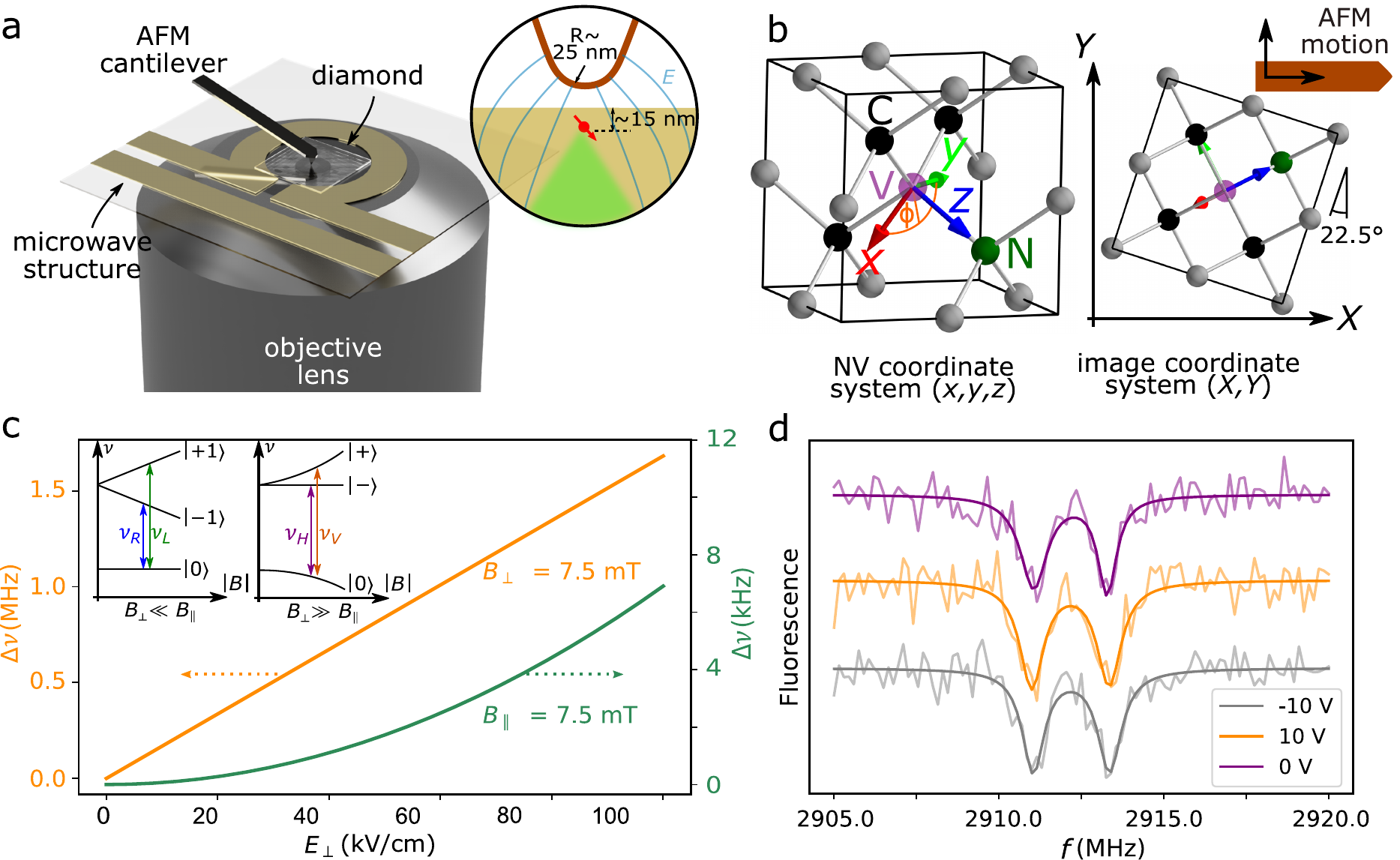}
	\caption{\textbf{a} Render of experimental apparatus showing the orientation of the confocal objective lens, diamond, AFM cantilever and microwave structure on a glass coverslip. Inset: Depiction of scanning AFM tip near a shallow NV (red arrow) center in diamond. \textbf{b} Orientation of the NV center in the diamond unit cell and in the imaging coordinates, $\phi$ represents both $\phi_B$ and $\phi_E$. The diamond sample has a $\langle 100 \rangle$ top surface and was placed down at an azimuthal misalignment of the diamond edges of 22.5\textdegree\, with respect to the imaging coordinates. \textbf{c} Simulation comparing the shift of the NV spin-resonances from an applied perpendicular electric field when in an aligned or perpendicular magnetic field. For the perpendicular magnetic field it can be seen that the electric field induces a much stronger and linear response. Inset: Spin-levels and resonances of NV center under magnetic fields aligned to be parallel or perpendicular to the NV axis. \textbf{d} ODMR spectra showing no discernible shift with an applied DC voltage, shown here are the \nfifteen\, hyperfine resonances of the upper spin branch ($\nu_+$) with an applied perpendicular magnetic field of $B_\perp = 8.7$ mT.}
	\label{figure-1}
\end{figure}

The nitrogen-vacancy (NV) center in diamond \cite{doherty2013nitrogen} is currently the only system capable of imaging elementary charges at room temperatures with nanoscale precision. Single defects have demonstrated impressive AC (DC) electric-field sensitivities of 202~$\text{V cm}^{-1}\text{ Hz}^{-1/2}$ (891~$\text{V cm}^{-1}\text{ Hz}^{-1/2}$) \cite{dolde2011electric} while ensemble measurements have achieved shot-noise limited AC sensitivities on the order of 1~$\text{V cm}^{-1}\text{ Hz}^{-1/2}$ \cite{PhysRevA.95.053417}. However, these results are limited to detection of macroscopic electric field features \cite{dolde2011electric, PhysRevA.95.053417} or charges internal to diamond \cite{dolde2014charge,PhysRevLett.121.246402}. In this work we greatly extend these capabilities by using a single NV center to perform nanoscale electric field imaging of an external source in ambient conditions. A shallow NV center is used to detect the vector components of the electric field produced by the charged tip of an atomic force microscope in contact mode with the diamond surface (figure \ref{figure-1}a). Imaging is then realized through nanoscale control of the AFM tip position relative to the NV center.

We achieve these results by exploiting the NV center's unique capabilities for quantum sensing. These include a bright optical fluorescence, a mechanism for optical spin initialization and readout, and the longest room-temperature coherence time for any solid state spin \cite{doherty2013nitrogen,Herbschleb2019}. The combination of these properties allow for single defect identification and high-fidelity measurement of individual spin-resonances. These properties have previously been applied for precision nano-magnetometry \cite{balasubramanian2008nanoscale,taylor2008high}, thermometry \cite{kucsko2013nanometre,neumann2013high,toyli2012measurement} and quantum computing \cite{childress_hanson_2013}. Electrometry is realized by measuring the Stark shift of the NV${}^-$ spin-triplet ground state through optically detected magnetic resonance\cite{dolde2011electric}.

The Hamiltonian describing the ground state spin of the NV center in the presence of magnetic ($\vec{B}$) and electric field ($\vec{E}$) is \cite{doherty2013nitrogen}
\begin{align}
	H = \left(D+d_\parallel E_z\right) \left(S_z^2-\frac{2}{3}\right) +\gamma_e\vec{B}\cdot\vec{S} +d_\perp\left[E_y\left(S_xS_y+S_yS_x\right) -E_x\left(S_x^2-S_y^2\right)\right],
	\label{eqn-hamiltonian}
\end{align}
where $D = 2.87$ GHz is the zero-field splitting \cite{loubser1977optical}, $d_\parallel = 0.35(2)$ Hz/(V/cm) and $d_\perp = 17(3)$ Hz/(V/cm) are the electric field susceptibility parameters \cite{van1990electric} in the axial and transverse NV directions, $\gamma_e = 28.0$ MHz/mT is the NV spin’s gyromagnetic ratio \cite{loubser1977optical}, and $\vec{S}$ are the NV spin-operators ($S=1$). Analysis of the Hamiltonian \eqref{eqn-hamiltonian} reveals that the electric field susceptibility is maximised when the magnetic field is aligned perpendicular to the NV axis \cite{1367-2630-16-6-063067}, as shown in figure \ref{figure-1}c. In such a case, the spin resonance frequencies due to the applied electric field are given by
\begin{align}
	\nu_\pm \approx \nu_\pm(0) + k_\parallel E_z \mp k_\perp E_\perp \cos(2\phi_B + \phi_E),
	\label{eqn-freq-shift}
\end{align}
where $\tan\phi_B = B_y/B_x$, $\tan\phi_E = E_y/E_x$, $B_\perp = \sqrt{B_x^2+B_y^2}$ and $E_\perp = \sqrt{E_x^2+E_y^2}$. Here $\nu_\pm(0)$ are the spin-resonances in the absence of the applied electric field. Uniquely, the NV doesn't sense a potential, charge, capacitance or an electric force like other methods but the measures the electric field directly. Furthermore, the electric field components and the electric field polar angle $\phi_E$ are immediately separable by varying $\phi_B$.

As shown in figure \ref{figure-1}d, we were unable to see a Stark shift of the spin-resonances when applying a DC voltage to the tip up to $\lvert V_\text{tip}\rvert \le 10$ V. This is consistent with existing theory that suggests mobile charges on the diamond surface causes significant screening of DC fields \cite{PhysRevApplied.14.014085}. In contrast, significant Stark shift were observed at AC frequencies even at low voltages ($\lvert V_\text{tip} \rvert <1$ V). This frequency dependence of the surface screening is not unsurprising, as the finite mobility of charges on the surface will limit screening at higher frequencies.

Due to the long time required to raster an image using single point measurements, it is desirable to have as fast as possible measurement time per pixel. For AC frequency measurements, a considerable increase in speed and signal-to-noise can be obtained using the so-called ‘Qdyne’ sensing method \cite{Schmitt2017submillihertz, degan2017frequency, Mizuno2020}. Here our spin-measurements perform like a lock-in amplifier, where we use a known reference signal which is phase referenced to our acquisition system. This reference signal could be the movement of the oscillating AFM cantilever or a time-varying voltage applied to the AFM cantilever. This is the first time Qdyne has been incorporated with scanning probe microscopy. In this work we always apply a time varying voltage to an AFM cantilever in contact mode which introduces a time-varying shift of the spin-resonance of $\Delta \nu_\pm(t) = \left(k_\parallel E_z \mp k_\perp E_\perp \cos(2\phi_B + \phi_E)\right)\sin(2\pi f t + \varphi)$. For imaging, we are interested in detecting the amplitude of this signal (and therefore $\vec{E}$) as a function of space.

\begin{figure}
	\centering
	\includegraphics[width=0.9\textwidth]{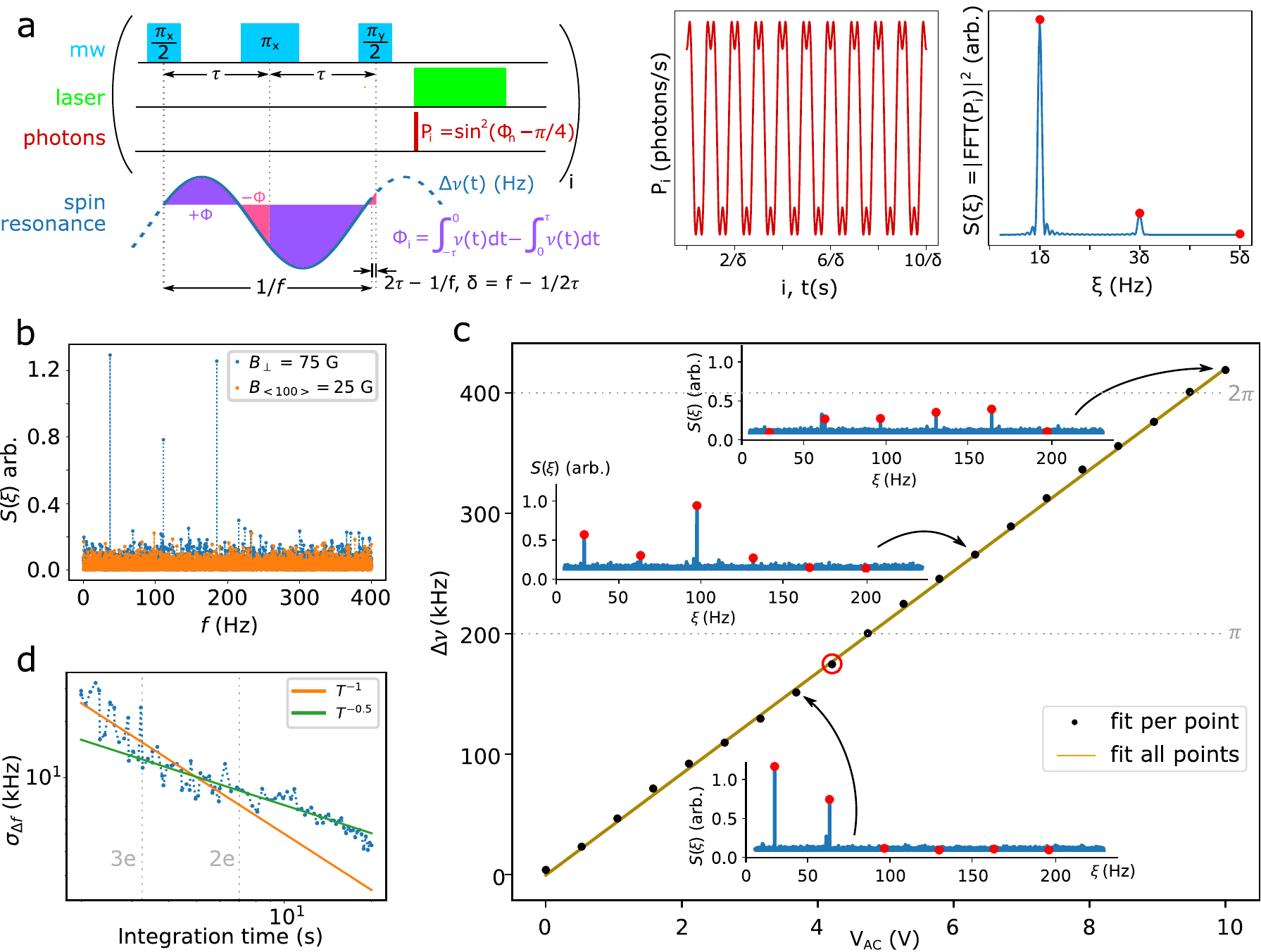}
	\caption{\textbf{a} Pulse sequence description for Qdyne measurement (see supplementary information). The detuning $\delta$ between the applied frequency $f$ and the inter-pulse spacing $\tau$ leads to a phase accumulation $\Phi_i$ that between oscillates across shots $i$ at the frequency $\delta$. Ultimately this results in sinusoidal fluorescence signal ($P_i$) oscillating at frequencies $(2n+1)\delta$, which can easily be analyzed using a power spectrum $S(\xi)=\lvert\text{FFT}(P_i(t))\rvert^2$. \textbf{b} Evidence of how the AC Stark shift is not apparent when the applied magnetic field is not perpendicularly aligned the NV axis, demonstrating that our AC signals are not due to some stray magnet field. Both measurements are for an applied AC voltage of amplitude 3.87 V, $\tau = 2\,\mu$s and $f = 250.037$ kHz, but with different applied magnetic fields. The microwave power was adjusted so the Rabi frequency (1/88 ns) was the same for each measurement. The non-perpendicular aligned field case clearly shows no signal resonances. \textbf{c} Example of linearization of data ($f=400.017$ kHz, $\delta=17$ Hz) from analyzing higher order resonances of signal spectra. The points are determined by fitting each resonance spectra independently and the line from fitting all resonance spectra simultaneously by optimizing a single coefficient that links $V_\text{AC}$ to $\Delta \nu$.  The horizontal dashed lines represent when the qubit vector has rotated around the equator of the Bloch sphere by $\pi$ during the phase accumulation. Insets show examples of spectra for specific data points. At larger voltages there are more higher order resonances present, resonances $(2n+1)\delta$ are indicated by the red dots.  \textbf{d} Standard deviation of extracted spin-frequency shift for increasing integration time. For the time scales used in imaging (max about 10 seconds per pixel) we do not reach the noise floor. We see a minimum signal noise of about 4 kHz. Data taken for this point is at 4.2 V (red circle in c). The vertical lines represent the amount of integration time to measure 3 and 2 elementary charges.}
	\label{figure-2}
\end{figure}

As shown in figure \ref{figure-2}a, by setting the time spacing ($\tau$) of our $\pi$-pulses to be slightly de-tuned from our applied signal frequency ($f$) we obtain a time-dependent fluorescence oscillating at a frequency given by that de-tuning ($\delta = f-1/2\tau$). If the signal amplitude produces large rotations of the qubit vector about the Bloch sphere ($\gtrsim \pi/4$) during the phase accumulation, then higher order odd harmonics ($\delta,3\delta,5\delta,\dots$) in the fluorescence signal become apparent. Qdyne has the benefit of mixing these harmonics down from the AC signal frequency $f \sim \mathcal{O}(\text{MHz})$ to the detuning frequency $\delta \sim \mathcal{O}(\text{Hz})$ making them substantially easier to observe.

Analysis of these harmonics allows us to greatly extend the dynamic range of our amplitude measurement, effectively unwrapping multiple rotations of the Bloch sphere. This is particularly important in imaging where the field magnitude near a source can vary sharply (e.g. $\sim1/r^2,~1/r^3$). Additionally, provided at least the second harmonic is present, this can be used to calibrate the signal magnitude using only the information from the signal itself. The specific amplitudes of the harmonics is dependent on the exact rotation of the qubit vector about the Bloch sphere. This self-calibration means that characterization of the optical emission, spin-contrast and other experimental parameters are not required.

An example of this extraction and linearization is shown in figure \ref{figure-2}c. Here we have used the power spectrum $S(\xi)$ to analyze the magnitude of the signal harmonics, their relationship with the applied signal amplitude is well defined and can be used to easily extracted the absolute frequency shift due to the applied electric field (see supplementary information for detailed derivation of the Qdyne sensing method). 

The standard deviation in the extracted frequency shift for increasing averaging time shows the normal $1/\sqrt{T}$ shot-noise dependence, we obtain an signal standard deviation of about 4 kHz after 11 seconds of averaging, giving us a frequency shift sensitivity of $\eta_{\Delta f} = $ \freqsensitivity\,kHz$/\sqrt{\text{Hz}}$. This value just represents the achievable signal given our noise, the real sensitivity to a source external to the diamond is dependent on the extent of the surface screening. Looking at the voltage applied to the AFM tip in figure \ref{figure-2}c shows a linear trend of 42 kHz frequency shift per volt, given our frequency shift sensitivity we obtain an AC voltage sensitivity of \voltagesensitivity\, V$/\sqrt{\text{Hz}}$. Bear in mind, this is strongly dependent on the location of the AFM tip relative to the NV, so it could be in principle improved. Modeling the AFM tip as a conducting sphere of radius 25 nm ($\approx$ AFM tip radius) with a single charge in the center, gives us an AC charge sensitivity of $\eta_e =\chargesensitivity$\, charges$/\sqrt{\text{Hz}}$. Therefore, we can detect the potential on the sphere equivalent to a single fundamental charge after 28 seconds of measurement. The longest measurement performed  was for 60 seconds, indicating that we measured the signal equivalent to 0.68\% of a single elementary charge. As expected, and as shown in figure \ref{figure-2}b we do not see an electrometry signal from an AC electric field when the magnetic field is not applied in a direction perpendicular to the NV axis.

\begin{figure}
	\centering
	\includegraphics[width=0.9\textwidth]{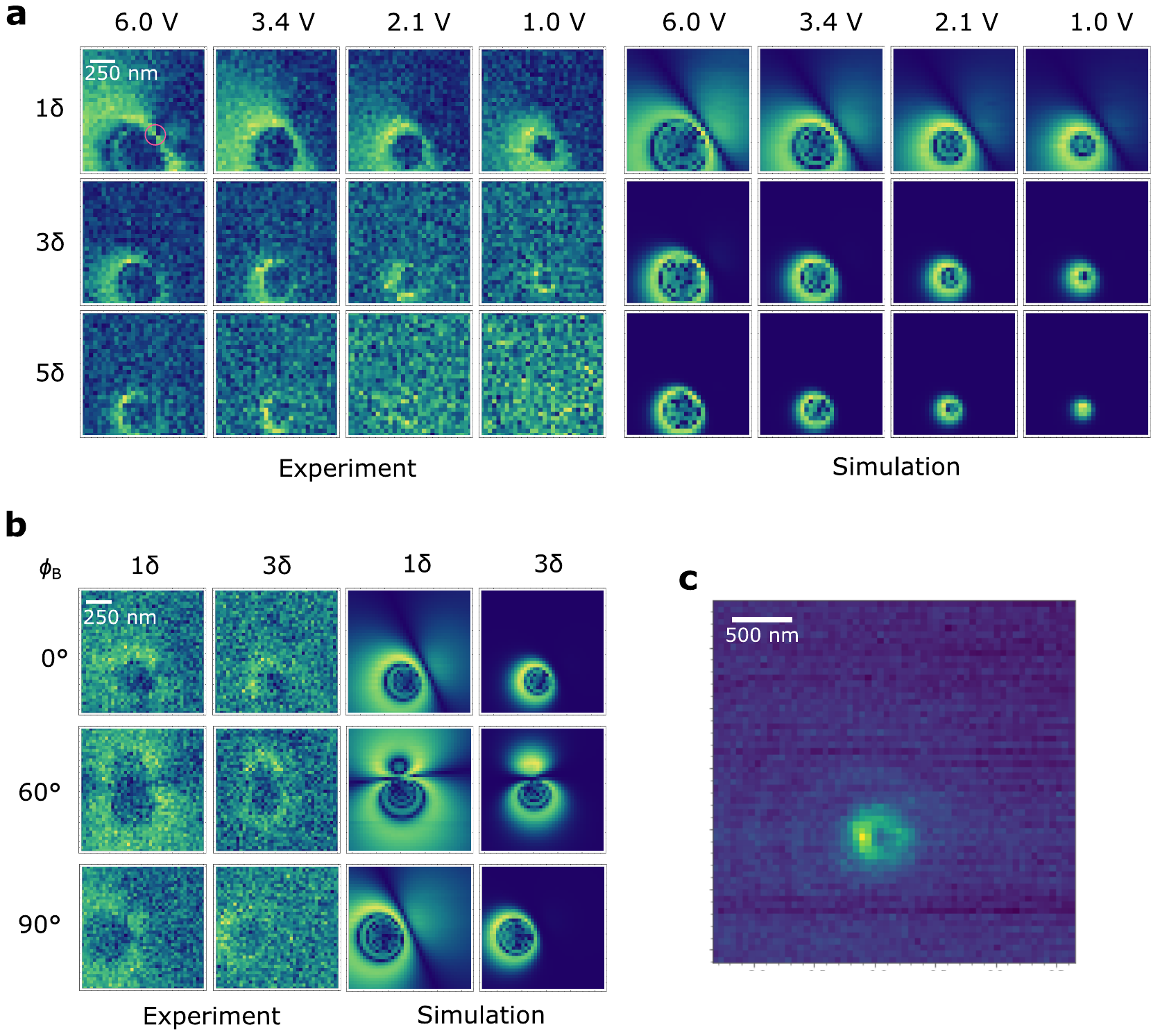}
	\caption{\textbf{a} Example of electric field imaging for varying applied voltages and for various signal resonances. Each pixel value is the integral of the power spectrum $S(\xi)$ across the signal resonance $(2k+1)\delta$. Here the applied field was at a frequency $f=250.017$ kHz. Each pixel was integrated for 3.3 s. The image is $30\times30$ pixels over a $1\mu$m square region, so each pixel is 33.3 nm in size. The upper left image has a circle denoting a region where a spatial resolution is pixel-size limited. \textbf{b} Dependence of rotating the polar magnetic field angle $\phi_B$. For each image a new $\phi_B$ magnetic field angle was chosen and the measurement re-ran. There are small shifts of the AFM position between images. \textbf{c} Example of near field fluorescence effects, you can see the dark quenching spot ($\sim 100$ nm diameter) in the center of the image when the AFM tip is directly above the NV and weaker set of surrounding rings demonstrating Purcell enhancement. The dark spot is used to find the AFM position relative to the NV and is used in the above simulations.}
	\label{figure-3}
\end{figure}

To perform imaging, the same measurement sequence is continuously applied whilst moving the position of the AFM tip relative to the NV in an XY raster motion. The same spectral analysis is then performed in a per pixel manner to extract electric field images. Figure \ref{figure-3} compares the electric field signal from various applied voltages for the first few spin-resonance harmonics to a COMSOL\textsuperscript{TM} simulation of the electric field from an AFM tip (see supplementary information). Note that this COMSOL\textsuperscript{TM} model is very well approximated by the field from a single monopole, as expected from a sharp point or a small conducting sphere at a set potential. We see very strong agreement of an image that corresponds to an NV at a distance 15 nm from the diamond surface and an AFM height of 30 nm above the surface, this height offset was probably from some contaminants on the surface of the diamond, most likely objective lens immersion oil. In this analysis we have ignored the effect of the electric field component $E_z$ as it contributes only about 2\% to the NV spin frequency shift. The lobed patterns and asymmetry in these images is due to the projection of the spherically symmetric monopole field from the AFM tip onto the plane transverse to the NV axis. The zero amplitude line that separates the two lobes is where $\cos\left(2\phi_B+\phi_E\right)$ passes through zero.

The effect of rotating the polar field angle $\phi_B$ is shown in figure  \ref{figure-3}b. Rotating $\phi_B$ adds an effective offset to $\phi_E$, similar to rotating the NV axis or diamond in the azimuthal direction, in the images this is seen as a rotation of the two lobes. This can be used as a handle to separate $\phi_E$ and the individual vector components of the electric field $E_x$ and $E_y$. We can also perform the reverse, and use a model of the electric field to extract the exact NV orientation (see supplementary information). We determine the exact orientation of the NV axes and ordering of the N and V along NV axis, as shown in figure \ref{figure-1}b. This is not possible without the application of an electric field \cite{1367-2630-16-6-063067}. The 0\textdegree\, and 90\textdegree\, images are similar because of the $\pi$ periodicity of $\phi_B$. However, when rotating the transverse field by 90\textdegree, the polarization of the spin-resonance flips (i.e. the lower/upper spin resonances flip from H/V polarized transitions), Since our microwaves were linearly polarized along only one direction, the 90\textdegree\, image was taken using the lower spin-resonance ($\ket{-}$) and 0\textdegree\, was taking using the upper spin-resonance ($\ket{+}$). In each case, the microwave power was adjusted such that the Rabi frequency matched between each image. Technically, using the other spin-resonance branch will flip the sign of the interaction in equation (\ref{eqn-freq-shift}), but since we are looking at the magnitude of the interaction only, this is not noticed.

The limit for the spatial resolution of this method has not been reached, any AFM or NV position drifts or deformities of the AFM tip geometry could limit our achievable resolution. However, we can see features in figure \ref{figure-3}a which are pixel-size limited putting an upper bound on our spatial resolution of 33.3 nm, which could be in principle improved by simply decreasing the image pixel size. 

Very close to the AFM tip ($\lesssim 100$ nm) we see quenching of the NV fluorescence (Fig. \ref{figure-3}c)), it is expected that this is a plasmonic quenching effect \cite{lakowicz2005radiative}. This could be alleviated by raising the AFM tip slightly off the surface of the diamond or changing to imaging a non-plasmonic sample. It does however provide a useful reference to find the position of the AFM tip relative the image axes. When the AFM tip is further from the NV, the NV fluorescence shows weak concentric rings of weak Purcell enhancement and quenching of the NV fluorescence. These effects are captured from the total NV fluorescence and are included into the analysis of the spin-signal images.

Further work will perform spectroscopy on charge trap mobility using the NV center in diamond to unlock information about the electrical properties of diamond surfaces. This understanding will allow us to limit surface screening to enhance NV electrometry and sensitivity. Additionally, electric field noise from the surface is suspected as a major factor in spin-decoherence for near-surface NV centers \cite{Kim2015decoherence-surface, PhysRevLett.118.197201}, greater understanding of the surface will be valuable in mitigating decoherence for shallow NV centers, vitally important for nanoscale imaging and sensing.

Due to the strong electric-field screening at DC we have only demonstrated sensing of AC signals\cite{PhysRevApplied.14.014085}. Limitations of our equipment restricted DC voltages that could be applied to our tip to $\pm10$~V. It has recently been demonstrated that DC screening effects can be circumvented at higher voltages\cite{bian2020nanoscale}, presumably through saturating charge traps local to the tip and NV. However, the sensitivity of such measurements is limited by the reduced coherence time of the NV center at DC. We instead propose imaging a DC electric field source (e.g. a static charge) through motion of the AFM cantilever (or diamond probe) to add an AC component. In this case the periodic cantilever motion could be used simultaneously as the Qdyne reference signal. We have tested this method with a mechanically oscillating magnetic AFM cantilever and confirmed that the phase/frequency of the AFM motion is stable enough to perform Qdyne (see supplementary information). 

In conclusion, we have demonstrated sensitive nanoscale imaging of the electric fields produced from a a voltage on an AFM tip very near a single NV. We see strong screening of the electric field at DC but significant AC fields. This is also the first research presented in the literature where AC electric fields have been sensed using an NV center. We describe and utilize the repetitive readout method `Qdyne' to greatly enhance our signal and imaging acquisition time and amplitude dynamic range, important for imaging a source with strong amplitude gradients. We demonstrate a charge sensitivity of $\eta_e =\chargesensitivity$ charges/$\sqrt{\text{ Hz}}$. We see strong agreement with resulting images and the model of the field from an AFM tip. In doing this we isolate the exact NV orientation and perform a demonstration of the vector imaging capability of this method. This work demonstrates that electric field vector imaging with single charge resolution is possible, and will pave the way for exciting new progress in measuring the fields from important nanoscale physical systems in ambient conditions. 

\section*{Methods}

The diamond sample is a 2$\times$2$\times$0.03 mm CVD electronic grade diamond membrane with a $\langle 100 \rangle$ surfaces that has undergone oxygen plasma surface termination, the NV we focused on was created by 10 keV implantation of nitrogen ions giving an estimated depth of 15 nm $\pm$ 5 nm \cite{pezzagna_NV_creation_efficiency}. We imaged though the diamond, so the NV centers are near the surface furthest from the objective lens and closest to the AFM. The sample was placed on a 0.17 mm thick glass cover slip with a Cu microwave waveguide deposited directly onto the glass. This was sufficient to achieve $> 10$ MHz electron spin Rabi frequencies. This sample had NV centers that generally exhibited electron spin coherence times $T_2 \lesssim 10~ \mu$s and $T_2^* \lesssim  1 \mu$s, but this deteriorated with the applied transverse magnetic field as spin-revivals from the nitrogen hyperfine field limits the usable electron spin evolution time to $\lesssim 5~ \mu$s for the field strengths used here. This effect is more pronounced for \nfifteen\, NV centers than for \nfourteen, as the zero-field splitting quadrupole moment of the \nfourteen\, is large compared to the off diagonal components of the hyperfine interaction \cite{Childress2006}. This effect is reduced for smaller magnetic field strengths, but large field strengths are desired to separate the $\ket{\pm}$ spin resonance by more than our Rabi frequency, in general we apply transverse magnetic field strengths $\lvert B_\perp\rvert > 70$ G. This problem could easily be circumvented by controlling the microwave polarization for selective spin resonance control.

Measurements were performed on an inverted confocal microscope/atomic force microscope (Asylum Research MFP3D-Bio). The confocal microscope had a scanning objective lens (Nikon TIRF 1.45/60x). The AFM has its own scanning sample stage where the diamond is placed, as such, the NV center in the diamond has to be followed by the confocal scanning stage as the AFM scanner rasters an image.
Excitation was with a 532 nm (Laser Quantum GEM) laser intensity modulated by a 200 MHz AOM, with gating provided by an RF switch. Fluorescence was coupled into a 25 $\mu$m core fiber and detected using a Excilitas single photon detector (SPCM-AQRH-14-FC). Microwaves were provided by a R\&S SMIQ03B signal generator with the XY phase component amplitude modulation from the internal I/Q modulator and amplified with a Minicircuits ZHL-16-43-S+ power amplifier. The electric field signal was provided from a Cr/Pt coated conducting AFM tip (Budget Sensors ElectriCont-G) in contact mode. The electric voltage signal was provided from a Stanford DS345 signal generator and connected via the MFP3D `chip’ crosspoint switch terminal. The timing of the photon arrivals was measured using a Swabian Instruments TimeTagger 20. Pulses controlling the AOM switch and XY microwave modulation was from a Swabian Instrument PulseStreamer 8/2. The TimeTagger, PulseStreamer, Stanford DS345 and SMIQ03B were all referenced to the SMIQ03B's internal 10 MHz reference clock. Vector magnetic field control is provided from a home build electromagnet based on a steel magnetic circuit arrangement placed around the confocal objective lens. This magnet is capable of generating $\pm 100$ G independently in each axis (XYZ).

\section*{Acknowledgments}

This work was supported in part by the Australian Research Council (ARC) and European Union research funding. M.W.D. acknowledges the ARC funding DP170102735. J.W. acknowledges funding from ASTERIQS and SMel (ERC grant number: 742610).

\section*{Contributions}

All experiments and data analysis was performed by M.S.J.B. Experimental apparatus was built by M.S.J.B. COMSOL$^\text{TM}$ modeling and analysis was performed by L.M.O. The diamond sample was provided by A.D. and J.W. Advice and guidance provided by L.P.McG, N.B.M. and project leadership from J.W. and M.W.D. All authors contributed to the manuscript with the initial draft written by M.S.J.B. and L.M.O.

\newpage

\newpage

\begin{thebibliography}{10}
	
	\bibitem{Yoo579}
	M.~J. Yoo, T.~A. Fulton, H.~F. Hess, R.~L. Willett, L.~N. Dunkleberger, R.~J.
	Chichester, L.~N. Pfeiffer, and K.~W. West.
	\newblock Scanning single-electron transistor microscopy: Imaging individual
	charges.
	\newblock {\em Science}, 276(5312):579--582, 1997.
	
	\bibitem{Martin2008}
	J.~Martin, N.~Akerman, G.~Ulbricht, T.~Lohmann, J.~H. Smet, K.~von Klitzing,
	and A.~Yacoby.
	\newblock Observation of electron-hole puddles in graphene using a scanning
	single-electron transistor.
	\newblock {\em Nature Physics}, 4(2):144--148, 2008.
	
	\bibitem{doi:10.1063/1.358819}
	Albert~K. Henning, Todd Hochwitz, James Slinkman, James Never, Steven Hoffmann,
	Phil Kaszuba, and Charles Daghlian.
	\newblock Two‐dimensional surface dopant profiling in silicon using scanning
	kelvin probe microscopy.
	\newblock {\em Journal of Applied Physics}, 77(5):1888--1896, 1995.
	
	\bibitem{doi:10.1063/1.102312}
	C.~C. Williams, J.~Slinkman, W.~P. Hough, and H.~K. Wickramasinghe.
	\newblock Lateral dopant profiling with 200 nm resolution by scanning
	capacitance microscopy.
	\newblock {\em Applied Physics Letters}, 55(16):1662--1664, 1989.
	
	\bibitem{Devoret2000}
	M.~H. Devoret and R.~J. Schoelkopf.
	\newblock Amplifying quantum signals with the single-electron transistor.
	\newblock {\em Nature}, 406(6799):1039--1046, 2000.
	
	\bibitem{PhysRevLett.65.3162}
	C.~Sch\"onenberger and S.~F. Alvarado.
	\newblock Observation of single charge carriers by force microscopy.
	\newblock {\em Phys. Rev. Lett.}, 65:3162--3164, Dec 1990.
	
	\bibitem{doi:10.1063/1.99224}
	Y. Martin, D.~W. Abraham, and H.~K. Wickramasinghe.
	\newblock High‐resolution capacitance measurement and potentiometry by force
	microscopy.
	\newblock {\em Applied Physics Letters}, 52(13):1103--1105, 1988.
	
	\bibitem{Cleland1998}
	A.~N. Cleland and M.~L. Roukes.
	\newblock A nanometre-scale mechanical electrometer.
	\newblock {\em Nature}, 392(6672):160--162, 1998.
	
	\bibitem{Bunch490}
	J.~Scott Bunch, Arend~M. van~der Zande, Scott~S. Verbridge, Ian~W. Frank,
	David~M. Tanenbaum, Jeevak~M. Parpia, Harold~G. Craighead, and Paul~L.
	McEuen.
	\newblock Electromechanical resonators from graphene sheets.
	\newblock {\em Science}, 315(5811):490--493, 2007.
	
	\bibitem{Salfi2010}
	J.~Salfi, I.~G. Savelyev, M.~Blumin, S.~V. Nair, and H.~E. Ruda.
	\newblock Direct observation of single-charge-detection capability of nanowire
	field-effect transistors.
	\newblock {\em Nature Nanotechnology}, 5(10):737--741, 2010.
	
	\bibitem{Lee_2008}
	J. Lee, Y. Zhu, and A. Seshia.
	\newblock Room temperature electrometry with {SUB}-10 electron charge
	resolution.
	\newblock {\em Journal of Micromechanics and Microengineering}, 18(2):025033,
	jan 2008.
	
	\bibitem{doi:10.1021/nn202852j}
	S. Ghatak, A.~N. Pal, and A. Ghosh.
	\newblock Nature of electronic states in atomically thin mos2 field-effect
	transistors.
	\newblock {\em ACS Nano}, 5(10):7707--7712, 2011.
	\newblock PMID: 21902203.
	
	\bibitem{Novoselov666}
	K.~S. Novoselov, A.~K. Geim, S.~V. Morozov, D.~Jiang, Y.~Zhang, S.~V. Dubonos,
	I.~V. Grigorieva, and A.~A. Firsov.
	\newblock Electric field effect in atomically thin carbon films.
	\newblock {\em Science}, 306(5696):666--669, 2004.
	
	\bibitem{10.1117/1.NPh.7.3.035002}
	L. Hanlon, V. Gautam, J.~D.~A. Wood, P. Reddy, M.~S.~J.
	Barson, M. Niihori, A.~R.~J. Silalahi, B. Corry, J. Wrachtrup,
	M.~J. Sellars, V.~R. Daria, P. Maletinsky, G.~J. Stuart,
	and M.~W. Doherty.
	\newblock {Diamond nanopillar arrays for quantum microscopy of neuronal
		signals}.
	\newblock {\em Neurophotonics}, 7(3):1 -- 12, 2020.
	
	\bibitem{doherty2013nitrogen}
	M.~W. Doherty, N.~B. Manson, P. Delaney, F. Jelezko, J.
	Wrachtrup, and L.~C.~L. Hollenberg.
	\newblock The nitrogen-vacancy colour centre in diamond.
	\newblock {\em Physics Reports}, 528(1):1--45, 2013.
	
	\bibitem{dolde2011electric}
	F.~Dolde, H.~Fedder, M.W.~Doherty, T.~N{\"o}bauer, F.~Rempp, G.~Balasubramanian,
	T.~Wolf, F.~Reinhard, L.C.L. Hollenberg, F.~Jelezko, and J~Wrachtrup.
	\newblock Electric-field sensing using single diamond spins.
	\newblock {\em Nature Physics}, 7(6):459--463, 2011.
	
	\bibitem{PhysRevA.95.053417}
	E.~H. Chen, H.~A. Clevenson, K.~A. Johnson, L.~M. Pham, D.~R.
	Englund, P.~R. Hemmer, and D.~A. Braje.
	\newblock High-sensitivity spin-based electrometry with an ensemble of
	nitrogen-vacancy centers in diamond.
	\newblock {\em Phys. Rev. A}, 95:053417, May 2017.
	
	\bibitem{dolde2014charge}
	F. Dolde, M.~W. Doherty, J. Michl, I. Jakobi, B. Naydenov,
	S. Pezzagna, J. Meijer, P. Neumann, F. Jelezko, N.~B.
	Manson, and J. Wrachtrup.
	\newblock Nanoscale detection of a single fundamental charge in ambient
	conditions using the $\mathrm{NV}{}^{\ensuremath{-}}$ center in diamond.
	\newblock {\em Phys. Rev. Lett.}, 112:097603, Mar 2014.
	
	\bibitem{PhysRevLett.121.246402}
	T.~Mittiga, S.~Hsieh, C.~Zu, B.~Kobrin, F.~Machado, P.~Bhattacharyya, N.~Z.
	Rui, A.~Jarmola, S.~Choi, D.~Budker, and N.~Y. Yao.
	\newblock Imaging the local charge environment of nitrogen-vacancy centers in
	diamond.
	\newblock {\em Phys. Rev. Lett.}, 121:246402, Dec 2018.
	
	\bibitem{Herbschleb2019}
	E.~D. Herbschleb, H.~Kato, Y.~Maruyama, T.~Danjo, T.~Makino, S.~Yamasaki,
	I.~Ohki, K.~Hayashi, H.~Morishita, M.~Fujiwara, and N.~Mizuochi.
	\newblock Ultra-long coherence times amongst room-temperature solid-state
	spins.
	\newblock {\em Nature Communications}, 10(1):3766, 2019.
	
	\bibitem{balasubramanian2008nanoscale}
	G.~Balasubramanian, IY~Chan, R.~Kolesov, M.~Al-Hmoud, J.~Tisler, C.~Shin,
	C.~Kim, A.~Wojcik, P.R. Hemmer, A.~Krueger, et~al.
	\newblock Nanoscale imaging magnetometry with diamond spins under ambient
	conditions.
	\newblock {\em Nature}, 455(7213):648--651, 2008.
	
	\bibitem{taylor2008high}
	J.M.~Taylor, P.~Cappellaro, L.~Childress, L.~Jiang, D.~Budker, PR~Hemmer,
	A.~Yacoby, R.~Walsworth, and M.D.~Lukin.
	\newblock High-sensitivity diamond magnetometer with nanoscale resolution.
	\newblock {\em Nature Physics}, 4(10):810--816, 2008.
	
	\bibitem{kucsko2013nanometre}
	G.~Kucsko, P.C.~Maurer, N.Y.~Yao, M.~Kubo, H.J.~Noh, P.K.~Lo, H.~Park, and M.D.~Lukin.
	\newblock Nanometre-scale thermometry in a living cell.
	\newblock {\em Nature}, 500(7460):54--58, 2013.
	
	\bibitem{neumann2013high}
	P. Neumann, I. Jakobi, F. Dolde, C. Burk, R. Reuter,
	G. Waldherr, J. Honert, T. Wolf, A. Brunner, J.H. Shim,
	et~al.
	\newblock High precision nano scale temperature sensing using single defects in diamond.
	\newblock {\em Nano letters}, 2013.
	
	\bibitem{toyli2012measurement}
	D.M.~Toyli, D.J.~Christle, A.~Alkauskas, B.B.~Buckley, C.G.~Van~de Walle, and
	D.D.~Awschalom.
	\newblock Measurement and control of single nitrogen-vacancy center spins above
	600 k.
	\newblock {\em Physical Review X}, 2(3):031001, 2012.
	
	\bibitem{childress_hanson_2013}
	L. Childress and R. Hanson.
	\newblock Diamond nv centers for quantum computing and quantum networks.
	\newblock {\em MRS Bulletin}, 38(2):134–138, 2013.
	
	\bibitem{loubser1977optical}
	J.H.N. Loubser and J.A.~Van~Wyk.
	\newblock Optical spin-polarisation in a triplet state in irradiated and
	annealed type 1b diamonds.
	\newblock {\em Diamond Research}, pages 11--14, 1977.
	
	\bibitem{van1990electric}
	E. Van~Oort and M. Glasbeek.
	\newblock Electric-field-induced modulation of spin echoes of nv centers in
	diamond.
	\newblock {\em Chemical Physics Letters}, 168(6):529--532, 1990.
	
	\bibitem{1367-2630-16-6-063067}
	M.~W. Doherty, J.~Michl, F.~Dolde, I.~Jakobi, P.~Neumann, N.~B. Manson, and
	J.~Wrachtrup.
	\newblock Measuring the defect structure orientation of a single nv$^{-1}$
	centre in diamond.
	\newblock {\em New Journal of Physics}, 16(6):063067, 2014.
	
	\bibitem{PhysRevApplied.14.014085}
	L.~M. Oberg, M.~O. de~Vries, L.~Hanlon, K.~Strazdins, M.~S.~J. Barson, M.~W. Doherty, and
	J.~Wrachtrup.
	\newblock {Solution to Electric Field Screening in Diamond Quantum
		Electrometers}.
	\newblock {\em Phys. Rev. Applied}, 14(1):14085, jul 2020.
	
	\bibitem{Schmitt2017submillihertz}
	S. Schmitt, T. Gefen, F.~M. St{\"u}rner, T. Unden, G. Wolff,
	C. M{\"u}ller, J. Scheuer, B. Naydenov, M. Markham,
	S. Pezzagna, J. Meijer, I. Schwarz, M. Plenio, A. Retzker,
	L.~P. McGuinness, and F. Jelezko.
	\newblock Submillihertz magnetic spectroscopy performed with a nanoscale
	quantum sensor.
	\newblock {\em Science}, 356(6340):832--837, 2017.
	
	\bibitem{degan2017frequency}
	J.~M. Boss, K.~S. Cujia, J.~Zopes, and C.~L. Degen.
	\newblock Quantum sensing with arbitrary frequency resolution.
	\newblock {\em Science}, 356(6340):837--840, 2017.
	
	\bibitem{Mizuno2020}
	K. Mizuno, H. Ishiwata, Y. Masuyama, T. Iwasaki, and M.
	Hatano.
	\newblock Simultaneous wide-field imaging of phase and magnitude of ac magnetic
	signal using diamond quantum magnetometry.
	\newblock {\em Scientific Reports}, 10(1):11611, 2020.
	
	\bibitem{lakowicz2005radiative}
	J.~R. Lakowicz.
	\newblock Radiative decay engineering 5: metal-enhanced fluorescence and
	plasmon emission.
	\newblock {\em Analytical biochemistry}, 337(2):171--194, 2005.
	
	\bibitem{Kim2015decoherence-surface}
	M.~Kim, H.~J. Mamin, M.~H. Sherwood, K.~Ohno, D.~D. Awschalom, and D.~Rugar.
	\newblock Decoherence of near-surface nitrogen-vacancy centers due to electric
	field noise.
	\newblock {\em Phys. Rev. Lett.}, 115:087602, Aug 2015.
	
	\bibitem{PhysRevLett.118.197201}
	B.~A. Myers, A.~Ariyaratne, and A.~C.~Bleszynski Jayich.
	\newblock Double-quantum spin-relaxation limits to coherence of near-surface
	nitrogen-vacancy centers.
	\newblock {\em Phys. Rev. Lett.}, 118:197201, May 2017.
	
	\bibitem{pezzagna_NV_creation_efficiency}
	S.~Pezzagna, B.~Naydenov, F.~Jelezko, J.~Wrachtrup, and J.~Meijer.
	\newblock Creation efficiency of nitrogen-vacancy centres in diamond.
	\newblock {\em New Journal of Physics}, 12(6):065017, 2010.
	
	\bibitem{Childress2006}
	L.~Childress, M.V.G. Dutt, JM~Taylor, AS~Zibrov, F.~Jelezko, J.~Wrachtrup,
	PR~Hemmer, and MD~Lukin.
	\newblock Coherent dynamics of coupled electron and nuclear spin qubits in
	diamond.
	\newblock {\em Science}, 314(5797):281--285, 2006.
	
\end{thebibliography}
\end{document}


\title{Supplementary Information: \\ Nanoscale vector electric field imaging using a single electron spin}
	\author{M.S.J Barson$^1$, L.M. Oberg$^1$, L.P. McGuinness$^1$, A. Denisenko$^2$, N.B. Manson$^1$, J. Wrachtrup$^2$,  \\ and M.W. Doherty$^{1*}$}
\date{$^1$ Laser Physics Centre, Research School of Physics, Australian National University, Acton, ACT 2601, Australia\\
	$^2$ 3. Physikalisches Institut, University of Stuttgart, Pfaffenwaldring 57, 70569 Stuttgart, Germany \\
	$^*$ Corresponding author. Email: marcus.doherty@anu.edu.au}
\maketitle

\section{Q-dyne}

Similar derivations are given in refs. \cite{degan2017frequency,Schmitt2017submillihertz}, but here we focus on extracting the amplitude of the signal not the frequency. In this method, as with standard spin-echo metrology, we begin by polarizing the spin in to the $\ket{0}$ state. We then apply a $\pi/2$-pulse which is resonant with a given spin-resonance, say $\ket{+}$. This creates a super-position state of the $\ket{0}$ and $\ket{+}$ spin-levels. We then let the spin-state precess around the Bloch sphere and accumulate a spin phase from the effect of the applied electric field. In our case, the applied electric field adds an AC perturbation on our spin-resonance frequency $\nu_+(t)=\nu_0 + A\sin(2\pi f t+\varphi)$, where $\nu_0$ is a DC component of spin-resonance frequency and $A$ and $f$ are the amplitude (in radians/s) and frequency (in Hz) of the electric field spin-resonance shift. Here, this frequency shift is $\nu_\pm(t) = \nu_0 + \left(k_\parallel E_z \mp k_\perp E_\perp \cos(2\phi_B + \phi_E)\right)\sin(2\pi f t + \varphi)$, therefore $A = \left(k_\parallel E_z \mp k_\perp E_\perp \cos(2\phi_B + \phi_E)\right)$ where the $k_{\parallel,\perp}$ terms must be converted from Hz/(V/cm) to radians/s/(V/cm) units. While in our case the electric field is the quantity of interest, we shall refer to a general $A$ for clarity in the subsequent discussion.on.

During the pulse sequence we apply a $\pi$-pulse every $\tau$ seconds. This flips the sign of the accumulated phase, allowing for destructive interference of the total accumulated phase at all frequencies not near $1/(2\tau)$. Let the measurement time for a single pulse sequence shot be denoted as $T_s$. Then the total accumulated phase beginning at a time $t$ can be calculated as
\begin{align}
	\Phi(t) = \int_t^{t+T_s}\nu_+(t)h(t)dt,
	\label{eqn-phi}
\end{align}
where $h(t)$ is a modulation function that describes the sign flipping effect of the $\pi-$pulses,
\begin{align}
	h(t) = \text{sgn}\left(\sin(2\pi t/2\tau)\right)).
\end{align}
The sign function is defined by
\begin{equation*}
  \text{sgn}(x) =
    \begin{cases}
      -1 & \text{if $x<0$}\\
      0 & \text{if $x=0$} \\
      1 & \text{if $x>0$},
    \end{cases}       
\end{equation*}
and therefore $h(t)$ is a square wave signal at a frequency $1/2\tau$.
\newline

We can expand the square wave $h(t)$ as the harmonic sum
\begin{align}
	h(t) = \frac{4}{\pi}\sum_{k=0}^{\infty}\frac{\sin\left(2\pi(2k+1)t/2\tau\right)}{2k+1}.
\end{align}
Now equation (\ref{eqn-phi}) can be written as
\begin{align}\label{bigPhi}
	\Phi(t) = \int_t^{t+T_s} \nu_0 h(t) dt + \frac{4A}{\pi}\int_t^{t+T_s}\sum_{k=0}^{\infty}\frac{\sin\left(2\pi(2k+1)t/2\tau\right)}{2k+1}\sin(2\pi f t+\varphi) dt.
\end{align}

Since the pulse sequence always has an even number of phase accumulation steps (i.e. $T_s = n\tau$, for even $n$) the DC component ($\nu_0$) cancels in the first integral of equation~\eqref{bigPhi}. The second integral can be expanded using the trigonometric product identity $2\sin(x)\sin(y) = \cos(x-y)-\cos(x+y)$ as
\begin{align}\label{bigPhi2}
	\Phi = \frac{2A}{\pi}\int_t^{t+T_s}\sum_{k=0}^{\infty}\frac{1}{2k+1}\left[\cos\left(2\pi\left(f-(2k+1)/2\tau\right)t+\varphi\right)-\cos\left(2\pi\left(f+(2k+1)/2\tau\right)t+\varphi\right)\right]dt.
\end{align}

Now consider when the applied frequency, $f$, is very close to that of our square wave modulation function frequency (i.e., $f=1/2\tau+\delta$, where $\delta \ll f$). Making this substitution in equation~\eqref{bigPhi2}, we find that only the term $\cos(2\pi\delta t)$ in the sum dominates the integral. By neglecting all faster terms $\mathcal{O}(f)$ or higher, we simplify as
\begin{align}
	\Phi &= \frac{2A}{\pi}\int_t^{t+T_s}\cos(2\pi \delta t+\varphi)dt \\
	&= \frac{2AT_s}{\pi}\left[\cos\left(2\pi\delta(t+T_s/2)+\varphi\right)\text{sinc}(\pi\delta T_s)\right].
\end{align}

The phase accumulated over the measurement time $T_s$ is usually limited by $T_2 \sim \mathcal{O}$($\mu$s) and $\delta \sim \mathcal{O}$(Hz). Then $\delta T_s \ll 1$ and the above simplifies to
\begin{align}
	\Phi = \frac{2AT_s}{\pi}\cos(2\pi\delta t + \varphi).
\end{align}
A final $\pi/2$ pulse about the $y$ axis rotates the state on the Bloch sphere and converts the accumulated phase into a spin-population, we project out the $\ket{0}$ component of this state using a laser pulse giving the following probability of measuring $\ket{0}$,
\begin{align}
	P_0 &= \sin^2\left(\Phi - \pi/4\right) \nonumber\\
	&= \frac{1}{2} \left(1+\sin(2\Phi)\right) \nonumber\\
	&= \frac{1}{2} \left(1+\sin(\frac{4AT_s}{\pi}\cos(2\pi\delta t + \varphi))\right)
	\label{eqn-P0-sincos}
\end{align}

We now ignore the DC component of equation \eqref{eqn-P0-sincos} as we are concerned explicitly with measuring time varying signals. Using the Jacobi-Anger expression to expand the composite of the $\sin$ and $\cos$ functions gives
\begin{align}
	P_0 = \sum_{k=0}^\infty J_{2k+1}\left(\frac{4AT_s}{\pi}\right)(-1)^k\cos\left((2k+1)(2\pi\delta t + \varphi)\right),
\end{align}
a sum of odd harmonic frequencies with their amplitudes given by Bessel functions of the first kind $J_{2k+1}$. Interestingly, if we chose to use a final $\pi/2$ pulse about the $x$ axis, the above expansion from equation (\ref{eqn-P0-sincos}) would be of the form $\cos(\cos())$ which would result in harmonics of even order ($2\delta,4\delta,\dots$). 

We now evaluate the Fourier transform of this signal for spectral analysis. Firstly, we include a description of the finite time of our measurement $T_m$ ($\sim\mathcal{O}$(s)) with the rect function
\begin{align}
	P_0 = \sum_{k=0}^\infty J_{2k+1}\left(\frac{4AT_s}{\pi}\right)(-1)^k\cos\left((2k+1)(2\pi\delta t + \varphi)\right)\text{rect}(t/T_m).
	\label{eqn-P0}
\end{align}
Consider that what we actually measure isn't $P_0$ directly, but a photon count rate
\begin{align}
	P = R(1-C(1-P_0)),
\end{align}
where $R$ is the photon count rate when in the $\ket{0}$ level ($\sim200$ kHz) and $C$ is the contrast between photon count rates of the levels $\ket{0}$ and $\ket{+}$, usually about 30\%. This introduces some additional pre-factors into our signal before the sum. Ignoring the components that don't change with $P_0$ yields $P \propto RCP_0$.

\begin{figure}[h!]
	\centering
	\includegraphics[width=0.6\textwidth]{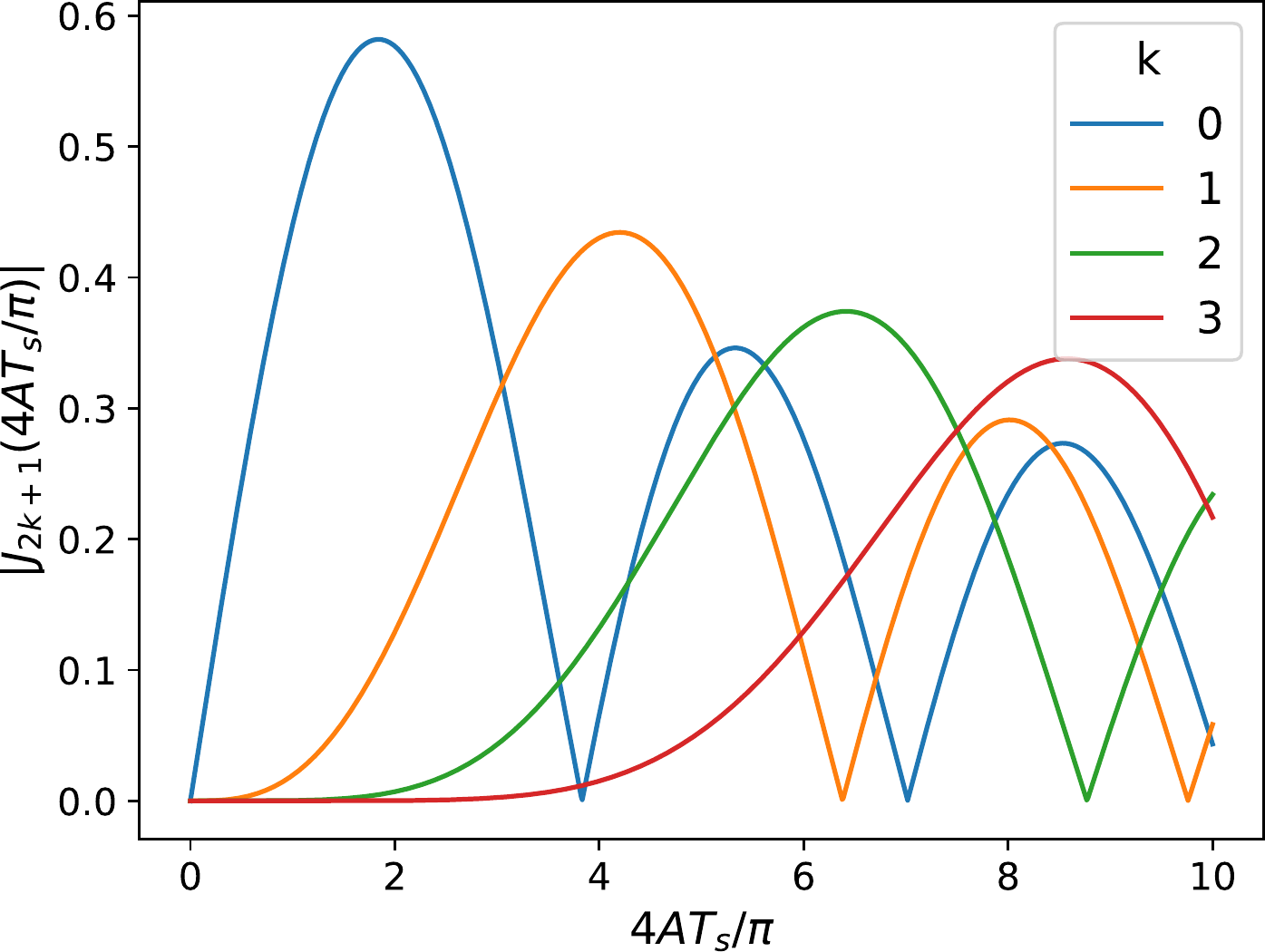}
	\captionof{figure}{Amplitudes of the first few resonances given by the Bessel function $J_{2k+1}$ for a given interaction strength $A$ (rad/s) and measurement time $T_s$ (s).}
	\label{figure-1}
\end{figure}

If we saturate the second harmonic resonance then we can use the relationship of the Bessel function amplitudes to get the absolute rotation of the qubit about the Bloch sphere (and determine $A$) without characterizing the parameters $R$ or $C$. More explicitly, the parameters $R$ and $C$ scale the values of figure \ref{figure-1} in the $y$ axis, but such scaling doesn't effect our ability to determine our amplitude $A$, represented in the $x$ axis. As such, we can ignore the effect of these parameters and not compromise our amplitude calibration.

Taking the Fourier transform of equation (\ref{eqn-P0}) gives
\begin{align}
	\mathcal{F}\left[P_0\right](\xi) = \sum_{k=0}^\infty J_{2k+1}\left(\frac{4AT_s}{\pi}\right)\frac{(-1)^kT_m}{2}e^{-2\pi i (2k+1)\varphi\xi}\left[\text{sinc}\left(T_m(\xi-(2k+1)\delta)\right)+\text{sinc}\left(T_m(\xi+(2k+1)\delta)\right)\right].
\end{align}

Ignoring the negative frequency components and measuring the power spectrum $S(\xi) = \lvert \mathcal{F}\left[P_0\right]\rvert^2$ gives,
\begin{align}
	S(\xi) = \sum_{k=0}^\infty \left[\frac{T_m}{2}J_{2k+1}\left(\frac{4AT_s}{\pi}\right)\right]^2\text{sinc}^2\left(T_m(\xi-(2k+1)\delta)\right).
\end{align}
When applying the square we can safely ignore cross terms in the sum. The spectral overlap of the $\text{sinc}\left(T_m(\xi-(2k+1)\delta)\right)$ terms for different $k=0,1,2,\dots$ is very small because the spectral width is much smaller than their spacing $1/T_m \ll 2\delta$.

To isolate the amplitude of the individual resonances we numerically integrate the experimentally obtained power spectrum across the harmonic resonances individually. The integrated power of the $k^\text{th}$ resonance is then
\begin{align}
	W_{2k+1} &= \int_{(2k+1)\delta-\epsilon}^{(2k+1)\delta+\epsilon} S(\xi) d\xi  \\
	&= \frac{\pi}{T_m}\left[\frac{T_m}{2}J_{2k+1}\left(\frac{4AT_s}{\pi}\right)\right]^2.
\end{align}
Here we have approximated that $\epsilon$ is large enough such that $\int_{-\epsilon}^{+\epsilon} \text{sinc}^2(T_m\xi)d\xi\approx \int_{-\infty}^{\infty} \text{sinc}^2(T_m \xi)d\xi = \pi/T_m$. This approximation is better than 3\% accurate for the measurement times $T_m$ we use here (usually  $T_m\sim50\times1/\delta\Rightarrow\epsilon\sim25\times\delta$). More importantly, this factor is also outside of the Bessel function, so as long as the treatment across resonances is consistent its effect on the extraction of the amplitude $A$ is suppressed.

Now we have an expression linking our experimental observable ($W_{2k+1}$) to each resonance in the signal and the total amplitude ($A$)
\begin{align}
	\left|J_{2k+1}\left(\frac{4AT_s}{\pi}\right)\right| = \sqrt{\frac{4W_{2k+1}}{\pi T_m}}.
	\label{eqn-bessel-W-relationship}
\end{align}
Note that Bessel functions are not invertible. Consequently, $A$ must be calculated numerically by fitting all resonances simultaneously. Given an array of harmonic amplitudes ($W_{2k+1}$) we can find a unique $A$ for a given $T_s$ that matches the harmonic pattern using equation (\ref{eqn-bessel-W-relationship}). This unique relationship allows us to perform calibration free absolute amplitude measurement.

\section{Static charge on an oscillating AFM tip}
\begin{figure}[h]
	\centering
	\includegraphics[width=0.8\textwidth]{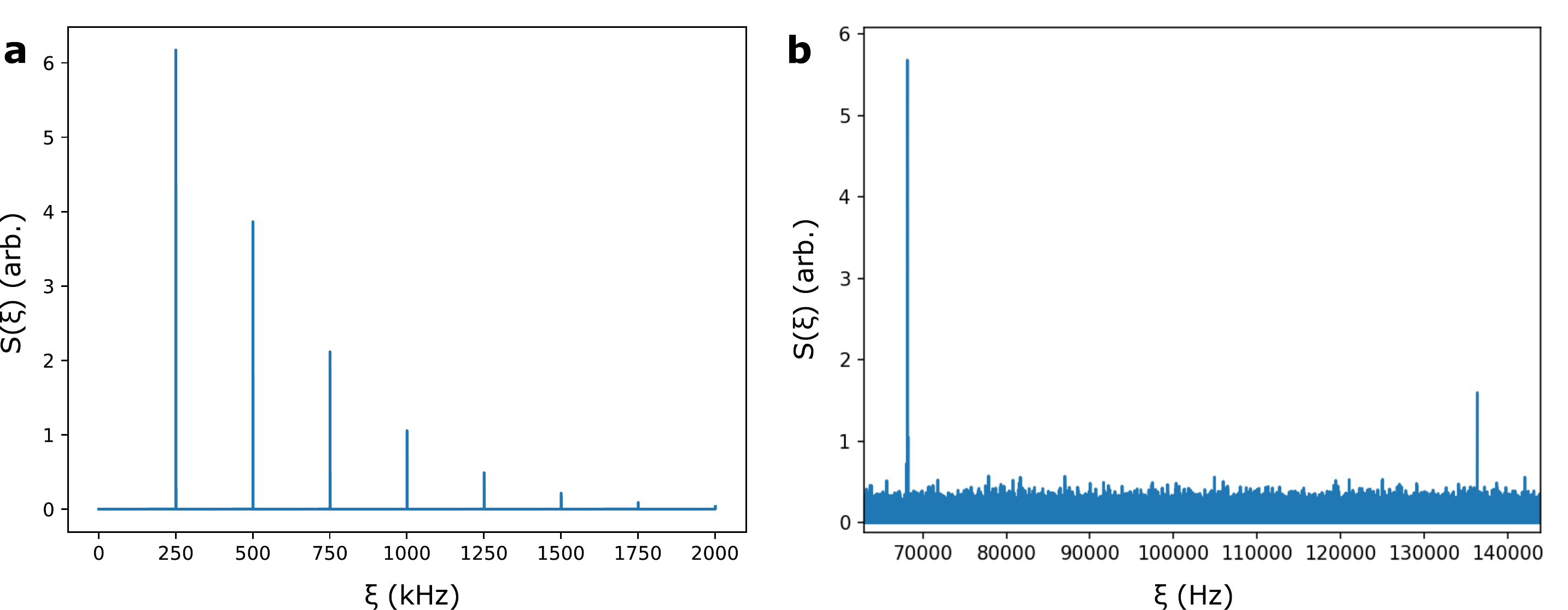}
	\captionof{figure}{\textbf{a} Power spectrum of the electric field $E_z$ as shown in equation \ref{eqn-Ez-mechanical} for a cantilever frequency of $f=250$ kHz. \textbf{b} Power spectrum from the spin-resonance shift from an oscillating magnetic AFM tip (MagneticMulti75-G) using a Ramsey style read-out. Measurement time was 10 seconds. The motion of the AFM cantilever was phase locked to the aqusition electronics.}
	\label{figure-3}
\end{figure}

If we have a static charge $q$ on the tip of an AFM cantilever undergoing AC oscillation in the $Z$ direction (such as with tapping mode) then our potential is
\begin{align}
	V(t) = k_d\frac{q}{\sqrt{X^2+Y^2+(Z-d_\text{AC}\sin(2\pi f t)-d_0)^2}},
\end{align}
where $d_\text{AC}$ is the amplitude of the cantilever oscillation and $d_0$ is its resting position. Which resuilts in the electric field components. 
\begin{align}
	E_x &= k_d \frac{qX}{\sqrt{X^2+Y^2+(Z-d_\text{AC}\sin(2\pi f t)-d_0)^2}} \\
	E_y &= k_d \frac{qY}{\sqrt{X^2+Y^2+(Z-d_\text{AC}\sin(2\pi f t)-d_0)^2}} \\
	E_z &= k_d \frac{q(Z-d_\text{AC}\sin(2\pi f t)-d_0)}{\sqrt{X^2+Y^2+(Z-d_\text{AC}\sin(2\pi f t)-d_0)^2}}
	\label{eqn-Ez-mechanical}
\end{align}
Although no longer sinusoidal this still produces an AC electric field that can be sensed using the same AC metrology techniques presented in the main text. The power spectrum $S(\xi)$ of $E_z$ is shown in figure \ref{figure-3}, and shows a much richer harmonic structure than a simple sinusoidal signal. However, this can still be easily extracted using the analysis presented in the previous section as the higher harmonics are suppressed by the filter function of the square wave $h(t)$.

Alternatively instead of applying the inverting $\pi-$pulse during the phase evolution we can simply perform a Ramsey ($\pi/2-\tau-\pi/2$) measurement per shot. This doesn't mix down the resulting signal to a much lower frequency but preserves harmonic structure of the original signal. The limitation here is that our phase accumulation period is limited to $T_2^*\sim\mathcal{O}(\lesssim 1\mu\text{s})$ as opposed to $T_2\sim\mathcal{O}(\lesssim 10\mu\text{s})$. In either case, we can easily use repetitive read-out techniques to extract the static charge on an oscillating cantilever. Allowing is to sense and AC signal from a DC charge, overcoming the DC surface screening described in the main text.

\section{Determining the exact NV orientation}

We use the standard definition of the NV coordinate system where the $z$ axis of the NV center is along the defect's symmetry axis from the V to the N. The $x$ axis of the NV center is then at 90\textdegree\, from $z$ towards a nearest neighbor carbon of the vacancy, that is, along a reflection plane of the defect's $C_{3V}$ symmetry. Finally the NV $y$ axis is simply orthogonal to the other two in a direction adhering to the right-hand rule. For an NV with a z direction along the $[111]$ crystal axis, x will be $[\bar{1}\bar{1}2]$ and y will be $[\bar{1}10]$. For an NV with a z direction along the $[\bar{1}\bar{1}\bar{1}]$ crystal axis, x will be $[11\bar{2}]$ and y will be $[1\bar{1}0]$.

\begin{figure}
	\centering
	\includegraphics[width=0.85\textwidth]{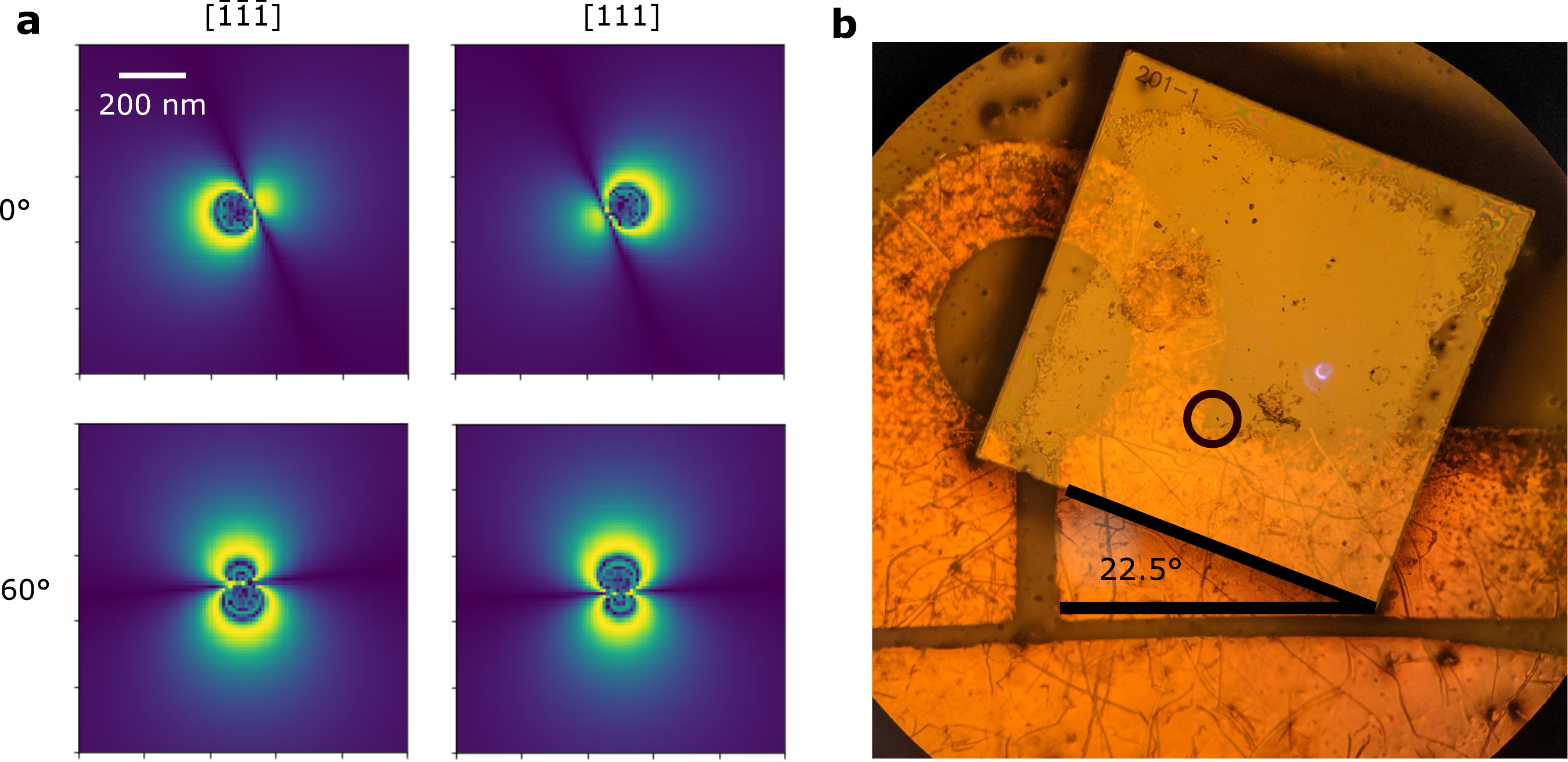}
	\captionof{figure}{\textbf{a} Comparison of the electric field pattern (for first resonance $\delta$) for two possible NV orientations $[\bar{1}\bar{1}\bar{1}]$ and $[111]$ that match our magnetic field charactersiation of the NV axis direction. We can clearly see that the asymmetry of the lobed patterns for the $[\bar{1}\bar{1}\bar{1}]$ matches the images shown in the main text. \textbf{b} Photograph of diamond roughly aligned to magnetic/image coordinates system showing the good agreement of the 22.5\textdegree\, azimuthal angle $\phi$ measured magnetically. This also tells us that this diamond has NV centres pointing towards the corners of the sample and therfore has $\langle 100 \rangle$ edges. The circle is roughly where the NV we imaged was located.}
	\label{figure-4}
\end{figure}

Our magnetic field when axially aligned with our NV center was along $(\theta,\phi) = (125$\textdegree$,\,22.5$\textdegree). The angle $\theta$ confirms this diamond sample has a $\langle 100 \rangle$ surface. The angle $\phi$ is from the placement of the diamond sample with edges not orthogonal to the magnetic field coordinate system. This is verified by the microscope image in figure \ref{figure-4}b. The NV axis is then either along a $[111]$
or $[\bar{1}\bar{1}\bar{1}]$ direction, where the sign ambiguity is because of the ordering of the N and V along that axis. Here we take the positive direction as being from V to N, as is convention. The other possible NV orientations (e.g. $[1\bar{1}1]$, $[\bar{1}1\bar{1}]$) are ruled out from the magnetic field alignment, i.e. we define $[110]$ in the $\phi = 22.5$ direction. Our first perpendicular magnetic field direction, called 0\textdegree\, in the main text is simply 90\textdegree\, shifted from the aligned direction $\theta \rightarrow \theta - 90$\textdegree, $(\theta,\phi) = (35$\textdegree$,\,22.5$\textdegree). If the NV was at a $[111]$ direction, this field would be the $x$ direction, if the NV was along a $[\bar{1}\bar{1}\bar{1}]$ this would be the $-x$ direction. Due to the $\pi$ periodic symmetry dependence of the spin-resonance shift on $\phi_B$ the $x,-x,y,-y$ magnetic field directions all produce the same field image so we can safely compare $[\bar{1}\bar{1}\bar{1}]$ and $[111]$ using the same perpendicular field direction. Using a simple monopole model for the electric field from the AFM (figure \ref{figure-4}a) for the two candidate NV directions we can see that $[\bar{1}\bar{1}\bar{1}]$ matches the images in the main text. To compare these two orientations we had to rotate the NV coordinate system for $[111]$ about $[001]$ by 22.5\textdegree\,$-$45\textdegree\, and for $[\bar{1}\bar{1}\bar{1}]$ by 22.5\textdegree\,$-$45\textdegree$+$180\textdegree\, such that the NV axial directions aligned with $\phi =$ 22.5\textdegree.

\section{COMSOL model of field from AFM tip}

\begin{figure}[h!]
	\centering
	\includegraphics[width=0.6\textwidth]{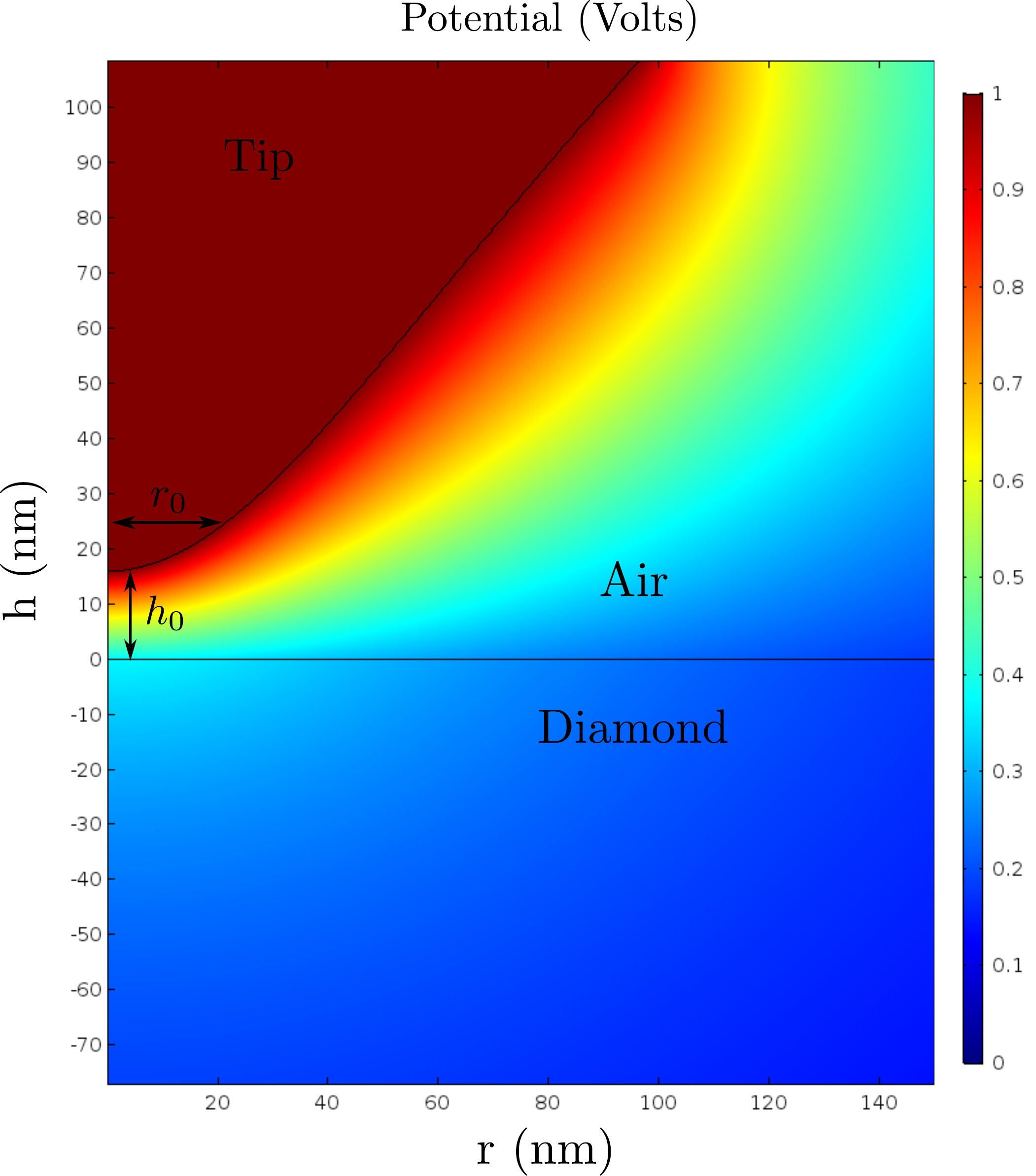}
	\captionof{figure}{COMSOL simulations demonstrating an example tip apex ($r_0=25$~nm, $h_0=15$~nm) and the resulting potential under application of 1 V. Note the use of axially-symmetric cylindrical coordinates.}
	\label{figure-COMSOL}
\end{figure}

\begin{figure}[h!]
    \centering
    \begin{subfigure}[b]{0.5\textwidth}
        \includegraphics[width=\textwidth]{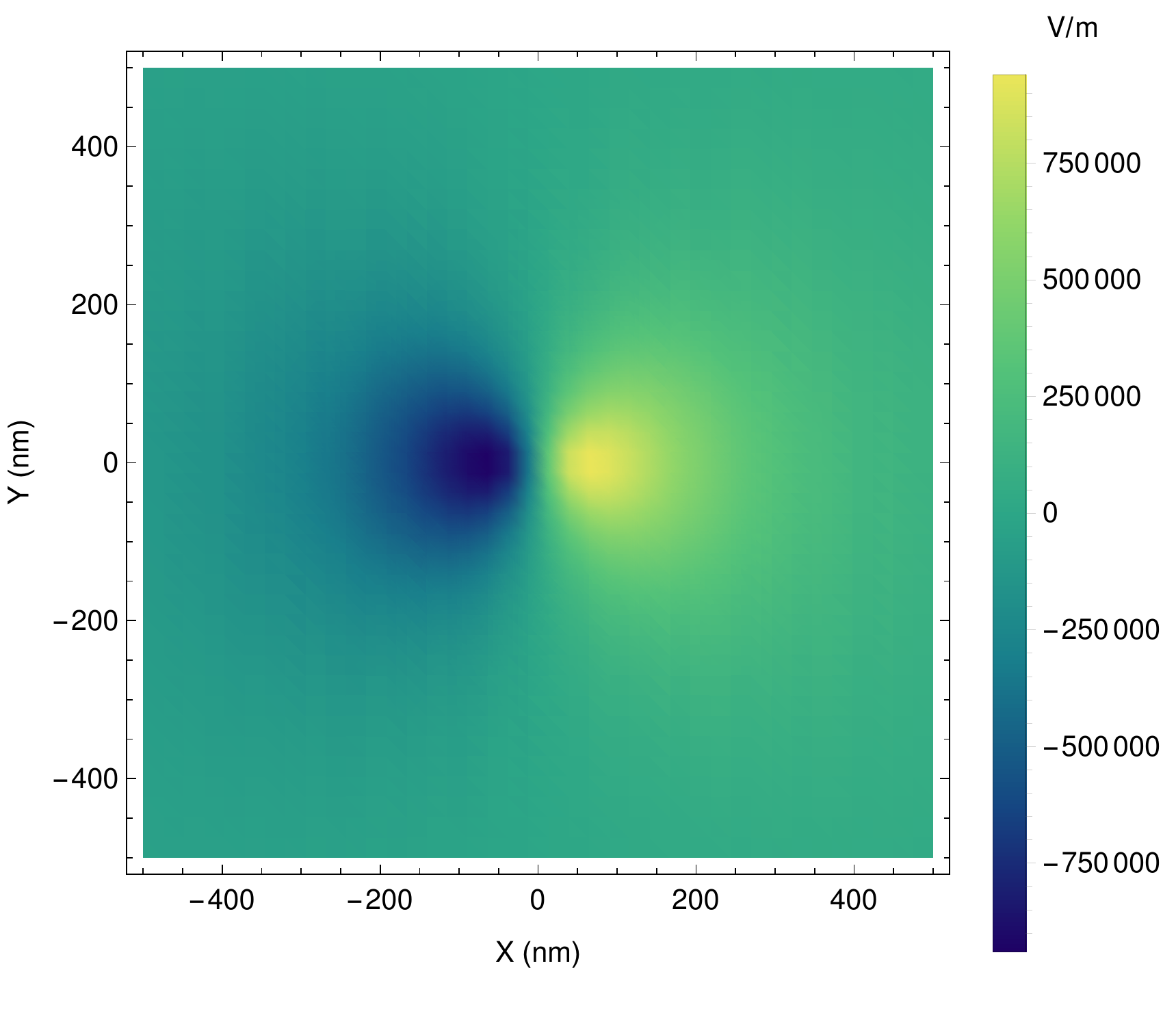}
        \caption{$E_x$.}
    \end{subfigure}
    ~
    \begin{subfigure}[b]{0.5\textwidth}
        \includegraphics[width=\textwidth]{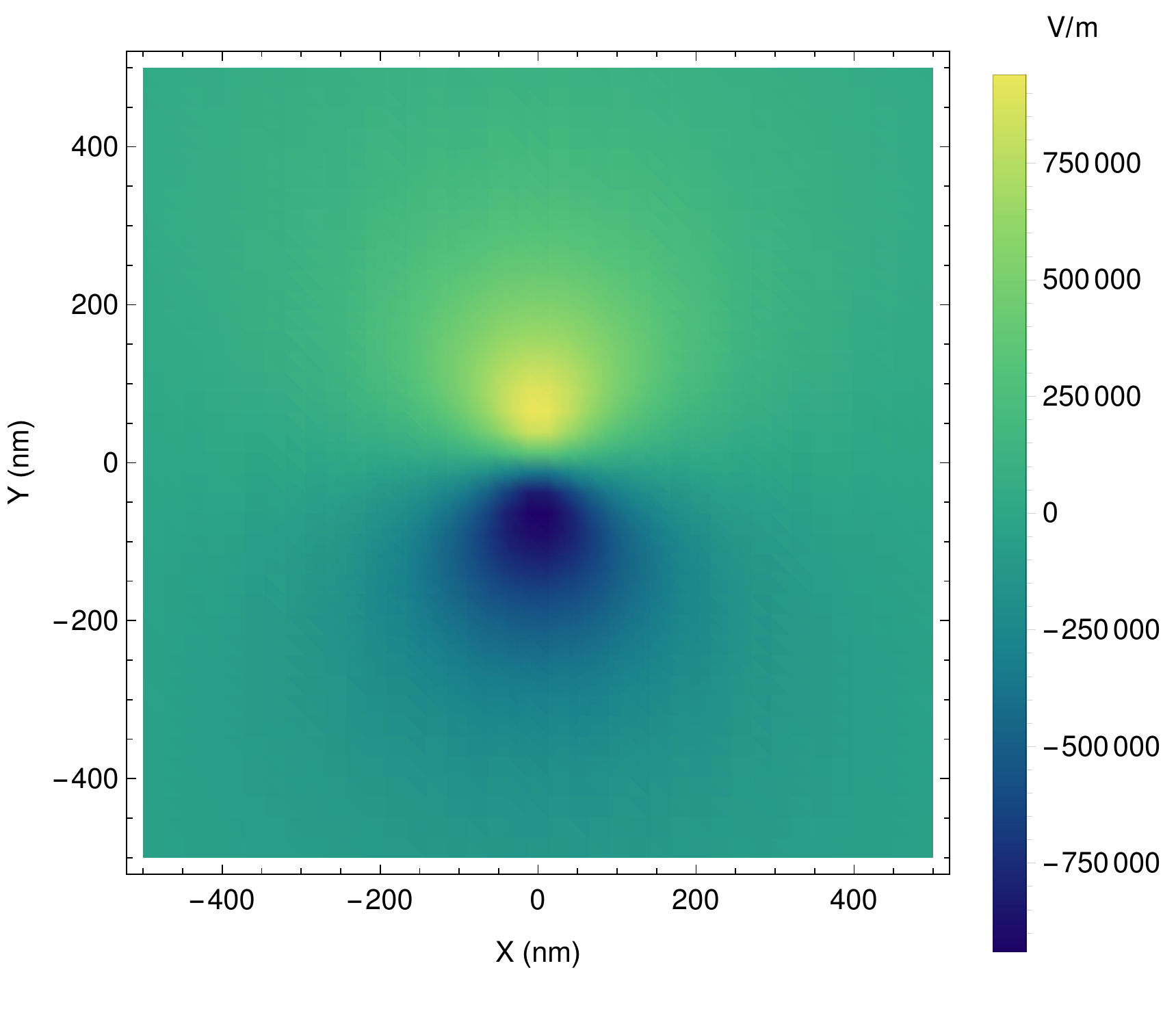}
        \caption{$E_y$.}
    \end{subfigure}
    ~
    \begin{subfigure}[b]{0.5\textwidth}
        \includegraphics[width=\textwidth]{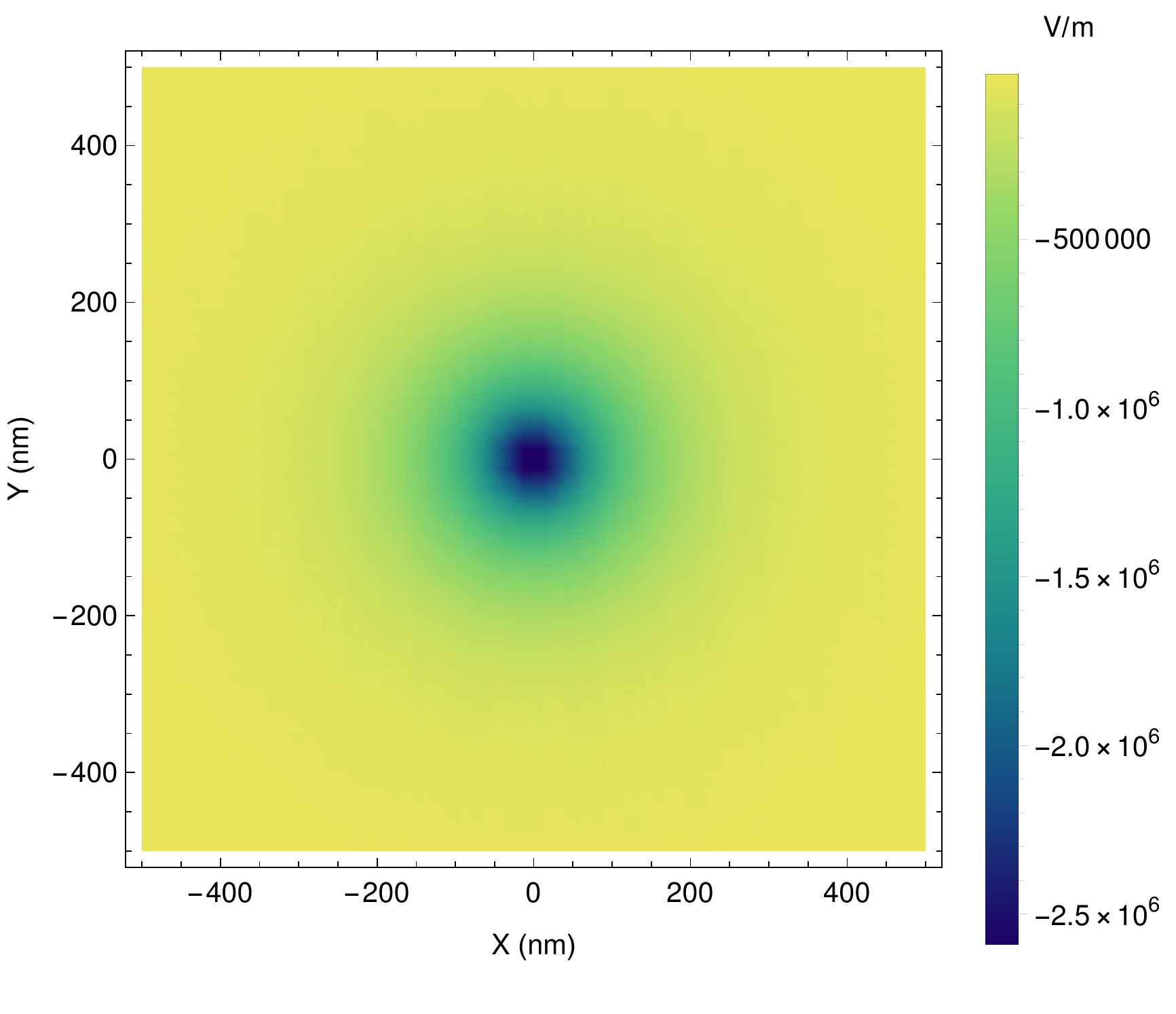}
        \caption{$E_z$.}
    \end{subfigure}
    \caption{COMSOL simulations of the electric field produced by a 1~V tip ($r_0=25$~nm, $h_0=15$~nm) sampled at 15~nm below the diamond surface.}
    \label{figure-field}
\end{figure}

COMSOL Multiphysics version 5.3 was used to simulate the electric field produced from a charged AFM tip above a diamond surface. The geometry of the tip was paramaterised in axially-symmetric cylindrical coordinates as $(h,r) = \left(h+h_0,h\tanh\left(\frac{r}{2r_0}\right)\right)$, where $h_0$ is the tip off-set above the diamond and $r_0$ represents an effective radius of the tip apex. This model provides an idealised approximation of the tip apex and is sufficiently accurate for the purposes of this work. A static potential and electric field was then calculated as a function of applied tip voltage. In Figure~\ref{figure-COMSOL} we present an example tip geometry ($r_0=25$~nm, $h_0=15$~nm) and the resulting potential when 1~V is applied to the tip. The vector components of the electric field have been depicted at a depth of 15~nm below the diamond surface in Figure~\ref{figure-field}. Note that these values depicted do not include the effects of surface screening. Following rotation into the NV reference frame these electric field values were used in conjunction with equation (2) of the main text to simulate the experimental results.

